\documentclass[11pt]{article}

\usepackage{authblk}

\usepackage{amsmath,amssymb,amsthm}
\usepackage{booktabs}
\usepackage{multirow}
\usepackage{makecell}
\usepackage{enumerate}
\usepackage{graphicx}
\graphicspath{{figs/}}
\usepackage{url}
\usepackage{color,subcaption}
\usepackage{geometry}
\usepackage{algorithm}
\usepackage{algpseudocode}
\usepackage{caption}
\usepackage[counterclockwise]{rotating}
\usepackage[T1]{fontenc}
\usepackage{setspace}

\usepackage{thmtools}
\usepackage{thm-restate}

\usepackage[bookmarksopen=false]{hyperref}

\usepackage[flushleft]{threeparttable}

\usepackage[round]{natbib}
\usepackage{tikz}
\newcommand{\widesim}[2][2]{
	\mathrel{\overset{#2}{\scalebox{#1}[1]{$\sim$}}}
}

\hypersetup{pdfborder = {0 0 0},colorlinks=true,linkcolor=blue,citecolor=blue}
\theoremstyle{definition}

\title{Statistical Machine Learning Meets High-Dimensional Spatiotemporal Challenges - A Case Study of COVID-19 Modeling}

\author[1]{Binbin Lin}
\author[2]{Yimin Dai} 
\author[1]{Lei Zou\thanks{Corresponding Author. \href{mailto: lzou@tamu.edu}{lzou@tamu.edu}}}
\author[2]{Ning Ning\thanks{Corresponding Author. \href{mailto: patning@tamu.edu}{patning@tamu.edu}}}
\affil[1]{Department of Geography, Texas A\&M University}
\affil[2]{Department of Statistics, Texas A\&M University}

\date{}

\begin{document}

\maketitle

\onehalfspacing

\begin{abstract}
\noindent 
Diverse non-pharmacological interventions (NPIs), serving as the primary approach for COVID-19 control prior to pharmaceutical interventions, showed heterogeneous spatiotemporal effects on pandemic management. Investigating the dynamic compounding impacts of NPIs on pandemic spread is imperative. However, the challenges posed by data availability of high-dimensional human behaviors and the complexity of modeling changing and interrelated factors are substantial. To address these challenges, this study analyzed social media data, COVID-19 case rates, Apple mobility data, and the stringency of stay-at-home policies in the United States throughout the year 2020, aiming to (1) uncover the spatiotemporal variations in NPIs during the COVID-19 pandemic utilizing geospatial big data; (2) develop a statistical machine learning model that incorporates spatiotemporal dependencies and temporal lag effects for the detection of relationships; (3) dissect the impacts of NPIs on the pandemic across space and time. Three indices were computed based on Twitter (currently known as X) data: the Negative and Positive Sentiments Adjusted by Demographics (N-SAD and P-SAD) and the Ratio Adjusted by Demographics (RAD), representing negative sentiment, positive sentiment, and public awareness of COVID-19, respectively. The Multivariate Bayesian Structural Time Series Time Lagged model (MBSTS-TL) was proposed to investigate the effects of NPIs, accounting for spatial dependencies and temporal lag effects. Results reveal a consistent lower level of public awareness about the pandemic in Louisiana. Twitter users in Vermont (Wyoming) consistently expressed a more optimistic (pessimistic) outlook regarding the pandemic. The developed MBSTS-TL model exhibited a high degree of accuracy. Determinants of COVID-19 health impacts transitioned from an emphasis on human mobility during the initial outbreak period to a combination of human mobility and stay-at-home policies during the rapid spread phase, and ultimately to the compound of human mobility, stay-at-home policies, and public awareness of COVID-19. The NPIs dataset offers a valuable resource for social scientists seeking to comprehend human activities during the pandemic. The MBSTS-TL model can be effectively applied to elucidate intricate interrelationships between human societies and infectious diseases. These findings furnish guidance for policymakers in implementing adaptive and phased strategies.		
\end{abstract}



\section{Introduction}

The COVID-19 pandemic, which emerged in 2020 and persisted for over three years, has profoundly impacted human society, posing significant threats to human health, disrupting social relationships, and devastating the economy \citep{li2020substantial,subramanian2021quantifying}. During the pandemic, governments worldwide implemented diverse policies to control the spread of the coronavirus. Meanwhile, individuals within different regions exhibited varying perceptions of the risks associated with COVID-19 and displayed divergent behaviors in response to the virus and adherence to relevant policies. These governmental and human responses, known as non-pharmacological interventions (NPIs), have played a crucial role in containing the pandemic, as evident in prior studies \citep{alharbi2020epidemiological, agusto2023impact, hadjidemetriou2020impact, kraemer2020effect, wellenius2021impacts, liu2021covid}. However, existing research has primarily focused on specific regions with similar policies or particular phases of the pandemic characterized by similar human responses. Thus, there is a pressing need to investigate the compounding effects of NPIs on the spread of COVID-19 and how these effects differ across space and evolve through time. The outcomes of such investigations are anticipated to inform the development of effective, localized policies and individual strategies for mitigating the adverse health consequences of future pandemics. 

There are two main challenges in modeling the compounding and evolving effects of NPIs on COVID-19’s health impacts: data availability and model complexity. Traditionally, data describing human responses to different events, e.g., the COVID-19 pandemic, can be obtained through surveys. However, conducting large-scale surveys during quarantine periods when face-to-face interactions are limited is challenging. Recent technological advances have provided new opportunities for monitoring human responses to the pandemic. Geospatial big data, e.g., web application data \citep{rovetta2020covid,tsao2021social} and sensor-based mobility data \citep{gao2020mapping,vinceti2020lockdown}, offer rich sources of information that can be used to delineate various dimensions of human behaviors, such as the strictness of COVID-19 related policies, public perceptions, and human mobility during the pandemic. These technological advances have the potential to provide new and valuable insights into the complex spatiotemporal dynamics of human responses and their impacts on the spread of pandemics. 

The second challenge is the model complexity. The impacts of diverse NPIs on COVID-19 are intricately intertwined \citep{vineis2003causality, galea2010causal}, evolving over time, unevenly dispersed across space, and spatiotemporal dependent \citep{li2022spatiotemporal}. These intricacies necessitate an advanced model capable of comprehending the high dimensional spatiotemporal data, incorporating the spatial dependence and time lag effects in the NPIs’ impacts, and effectively capturing the spatiotemporal varied effects of specific NPIs on COVID-19 spread. Conventional statistical models face challenges in handling high-dimensional data and uncovering interconnected relationships due to their limited capacity to capture complex patterns in such data. On the other hand, deep learning models, with their vast number of parameters, are considered "black boxes" because they lack interpretability in explaining relationships among variables. Despite their differences, both model types are susceptible to over-fitting during construction, wherein they may excessively fit the training data and fail to generalize well to new, unseen data, compromising their predictive performance. The emergence of statistical machine learning models \citep{sugiyama2015introduction} facilitates the ability to interpret relationships based on data-driven approaches, providing opportunities to model the complex impacts of NPIs on COVID-19 spread accurately.

This study analyzed the NPIs data, i.e., stay-at-home policies, public awareness and sentiment toward COVID-19, and human mobility, as well as the COVID-19 health impacts data in the U.S. at the state level. The year 2020 was chosen as the study period, considering the release of the first COVID-19 vaccination on December 14, 2020, and its considerable influence on the relationships between NPIs and the COVID-19 spread. The objectives of this study are threefold: (1) to reveal the spatiotemporal varied NPIs during the COVID-19 pandemic using data from social media, web applications, and smartphone sensors; (2) to develop a statistical machine learning model that incorporates spatiotemporal dependence and time lag effects for relationship detection; (3) to unravel the impacts of NPIs on the pandemic's health outcomes across time and space. The overarching hypothesis posits that stay-at-home policies and human mobility have a greater contribution to controlling the spread of COVID-19 compared to public awareness and sentiment. The knowledge gained from this study could provide an insightful understanding of NPIs’ impacts on COVID-19 spread and inform decision-making and policymaking for pandemic control. The developed framework can be used to model the relationships between other high-dimensional human responses and infectious diseases. 

\section{Background}

\subsection{Spatiotemporal COVID-19 modeling}

The COVID-19 pandemic has had an unequal impact on different regions and populations throughout its progression. To mitigate the uneven effects of the pandemic, it is crucial to understand how it spreads across different locations and over time. Spatiotemporal modeling of COVID-19 has emerged as a critical tool for achieving a detailed and accurate understanding of disease transmission patterns. 

Epidemiological models are crucial in modeling and predicting the spread of COVID-19, with the widely used Susceptible-Infected-Recovered (SIR) compartment model \citep{cooper2020sir,chen2020time,wangping2020extended}. The SIR model divides a population into susceptible, infected, and recovered compartments. Through an ordinary differential equation (ODE) system to describe the dynamics and flows between the compartments, the SIR model can estimate important metrics of a pandemic such as the basic reproduction number ($R_{\mathrm{0}}$) \citep{altmann1995susceptible}. However, this model assumes homogeneous dynamics across geographic areas, which does not reflect the spatial heterogeneity of the pandemic's spread and impacts. To address this limitation, researchers have developed spatial SIR-type models that incorporate spatial interactions between locations. For example, \cite{hatami2022simulating} developed a spatial Susceptible-Exposed-Infectious-Recovered (SEIR) model, incorporating a distance model describing pairwise relationships between studied locations with a traditional SEIR model. Another study by \cite{hou2021intracounty} developed human mobility flow-augmented stochastic SEIR model, applying an unsupervised machine learning algorithm to partition a county into multiple distinct subregions based on observed human mobility flow data. \cite{ionides2023iterated} and \cite{ning2023iterated} considered metapopulation systems characterized by strong dependence through time within a single unit and relatively weak interactions between units.

Spatial statistical regression models offer another approach to modeling the dynamics of epidemic spread. The Geographically and Temporally Weighted Regression (GTWR) model \citep{fotheringham2015geographical} is an extension of the Geographically Weighted Regression (GWR) model that incorporates temporal weighting. By allowing for the identification of local variations in the relationship between predictor variables and the response variable over time and space, the GTWR model has been successfully applied to estimate and forecast the spatial and temporal dynamics of COVID-19 spread in various regions \citep{chen2021modeling,liu2020space}. Similarly, the spatial error model \citep{wong2020spreading}, spatial lag model \citep{hafner2020spread}, and spatial vector autoregression model \citep{xie2015spatial} have been widely used to account for the effects of spatially structured errors and spatial dependence in modeling the spread of COVID-19. Researchers have also developed novel methods such as Bayesian network-based spatial predictive modeling \citep{dlamini2022bayesian} and the structured Gaussian process (SGP) model \citep{ak2022spatial} to incorporate spatial and temporal features into COVID-19 prediction. Specifically, \cite{dlamini2022bayesian} conducted Bayesian network-based spatial predictive modeling to delineate the dynamics of COVID-19 spread. This model considers proximity referral health facilities, churches, and shopping facilities as spatial variables. Incorporating spatial variables with daily traffic data and the proportion of youth, this model effectively identified COVID-19's potential geographic spread and the underlying influencing factors in Eswatini. \cite{ak2022spatial} constructed an SGP model integrating spatial (geographical coordinates and location-specific demographic information) and temporal features (the day, month, and year information of the reported case counts) to forecast the outbreak of COVID-19. 

\subsection{Governmental and Public Responses to COVID-19}

The COVID-19 pandemic has been acknowledged as a global crisis by the World Health Organization \citep{WHO2022url}. With humans serving as carriers and playing a crucial role in the dissemination of the disease, the importance of human responses to COVID-19 cannot be downplayed. Given the swift worldwide transmission of COVID-19 and the absence of an effective vaccine or treatment for this newly emerged infectious disease during its first outbreak, NPIs have emerged as one of the primary strategies to mitigate the spread of COVID-19 in 2020. 

Public health policies were emphasized as crucial NPIs to mitigate the rapid transmission of COVID-19. In China, for example, the government implemented the zero-COVID-19 policy until December 2022, employing large-scale testing, contact tracing technology, nationwide mask-wearing, and mandatory isolation of infected individuals to control the pandemic \citep{burki2020china}. In the U.S., California became the first state to enforce a stay-at-home or shelter-in-place order \citep{Newsom2020url}. In March 2020, the New York City public school system, the largest in the U.S. with 1.1 million students, shut down, while Ohio mandated the closure of restaurants and bars \citep{CDC2020url}. To assess the effectiveness of COVID-19 control policy measures, \cite{liu2021impact} conducted a study using panel regression to estimate the impact of 13 categories of COVID-19-related policies on reducing transmission across 130 countries from January to June 2020. Their findings revealed a strong positive correlation between strict policies such as school closures and internal movement restrictions and a decreased reproduction number of COVID-19. Another study conducted by \cite{dainton2021quantifying} examined the effects of COVID-19 lockdown policies on changes in human mobility, utilizing Google Mobility data from five contiguous public health units in the Greater Toronto Area  in Ontario, Canada, between March 1, 2020, and March 19, 2021. The study assessed the subsequent impact of changes in human mobility on the effective reproduction number of COVID-19, $R_0$, using Pearson correlation. The results indicated that, with enhanced lockdown measures, human mobility in York decreased significantly, particularly in retail, transit stations, and workplaces, leading to a reduction in $R_0$ after 14 days.

Additionally, the public perception of COVID-19 varied among residents in different regions or stages of the pandemic. This resulted in uneven adherence to recommended policies and personal protective behaviors, such as wearing masks and practicing proper hand hygiene, ultimately leading to distinct spatiotemporal patterns of COVID-19 transmission. \cite{cinarka2021relationship} conducted a study using Google search volumes for COVID-19 symptoms as indicators of public awareness in Turkey, Italy, Spain, France, and the United Kingdom. The dynamic conditional correlation analysis method was employed to explore the relationships between Google search volumes and the COVID-19 spread. The findings revealed that the Google search volumes for symptoms such as fever, cough, and dyspnea were closely correlated with new COVID-19 cases during the initial outbreak of the pandemic. \cite{jun2021impact} utilized Google's relative search volume (RSV) as an indicator of public awareness regarding COVID-19 and employed a vector autoregression model to investigate its association with new COVID-19 cases in 37 countries in the Organization for Economic Cooperation and Development (OECD). The results demonstrated that increased public awareness was associated with a heightened interest in COVID-19 testing, ultimately aiding in the early detection of new cases. \cite{agusto2023impact} employed ordinary differential equations to estimate the impact of public sentiment on the spread of COVID-19 in Australia, Brazil, Italy, South Africa, the United Kingdom, and the U.S. between January and June 2020. Public sentiments (both positive and negative) were evaluated using COVID-19-related tweets from Twitter. The findings indicated that positive public sentiments were associated with a reduction in disease burden within the community.

Nonetheless, current research has predominantly concentrated on specific regions with similar policies or particular phases of the pandemic characterized by comparable human responses. Consequently, there is an urgent necessity to examine the discrepancies in the cumulative impact of NPIs on the transmission of COVID-19, along with how these effects vary across different geographical locations and evolve over time.

\subsection{High-Dimensional Spatiotemporal Statistical Modeling}

High-dimensional spatiotemporal challenges arise when dealing with data that involve both space and time, and where there are a large number of variables, locations, and time points. In recent years, there have been several statistical advances in addressing these challenges. 

One direction is to use regularization methods such as Lasso \citep{tibshirani1996regression} or Elastic Net \citep{zou2005regularization}, which can help reduce the number of variables by assigning small coefficients to irrelevant variables. These methods can also identify important variables and their interactions. Another solution is dimension reduction which aims to reduce the number of variables in the data, while still capturing the relevant information. Principal component analysis \citep{zou2006sparse, ning2021spike}, factor analysis \citep{bhattacharya11,pati14}, and wavelet-based methods \citep{clyde1998multiple, huang2014nonlinear} are examples of dimension reduction techniques that have been applied to spatiotemporal data. These methodological advancements have substantially contributed to addressing the difficulties associated with analyzing high-dimensional and spatiotemporal data. As a result, there have been improvements in model development and predictions to resolve questions across various disciplines, such as environmental science, epidemiology, and climate modeling. Nonetheless, methodologies capable of simultaneously managing high-dimensional issues and spatiotemporal data remain scarce, primarily due to the intricate nature of feature selection tasks within complex structures.

The advances in the Bayesian structural time series (BSTS) model bring opportunities to address the challenges in high dimensional spatiotemporal statistical modeling.  BSTS \citep{scott2014predicting} is a statistical technique used to select features, forecast temporal trends, and infer causal impacts \citep{brodersen2015inferring}. The model is designed to work with time series data by incorporating various components, such as seasonality, trends, or auto-regression. It can also accommodate external regressors, which makes it possible to perform inferences about the impact of regressors on the response. Based on BSTS, the Multivariate BSTS (MBSTS) \citep{qiu2018multivariate,ning2021mbsts} has been proposed as a novel tool for inferring and predicting multiple correlated time series. MBSTS can select features from a pool of contemporary predictors while simultaneously training models for each time series, which reduces over-fitting and eliminates unessential or misleading predictors. In other words,  MBSTS can choose distinct predictor sets for each target time series for each Markov chain Monte Carlo (MCMC) iteration from high-dimensional data. 


In the context of human-pandemic systems, high-dimensional NPIs exhibit intricate inter-dependencies and have compounded impacts on epidemics. In this scenario, the MBSTS model, due to its inherent capability for feature selection and over-fitting prevention, is suitably employed to detect the compounded effects of NPIs on pandemic transmission. 

\section{Data}

\begin{figure}[!thbp]
    \centering
    \includegraphics{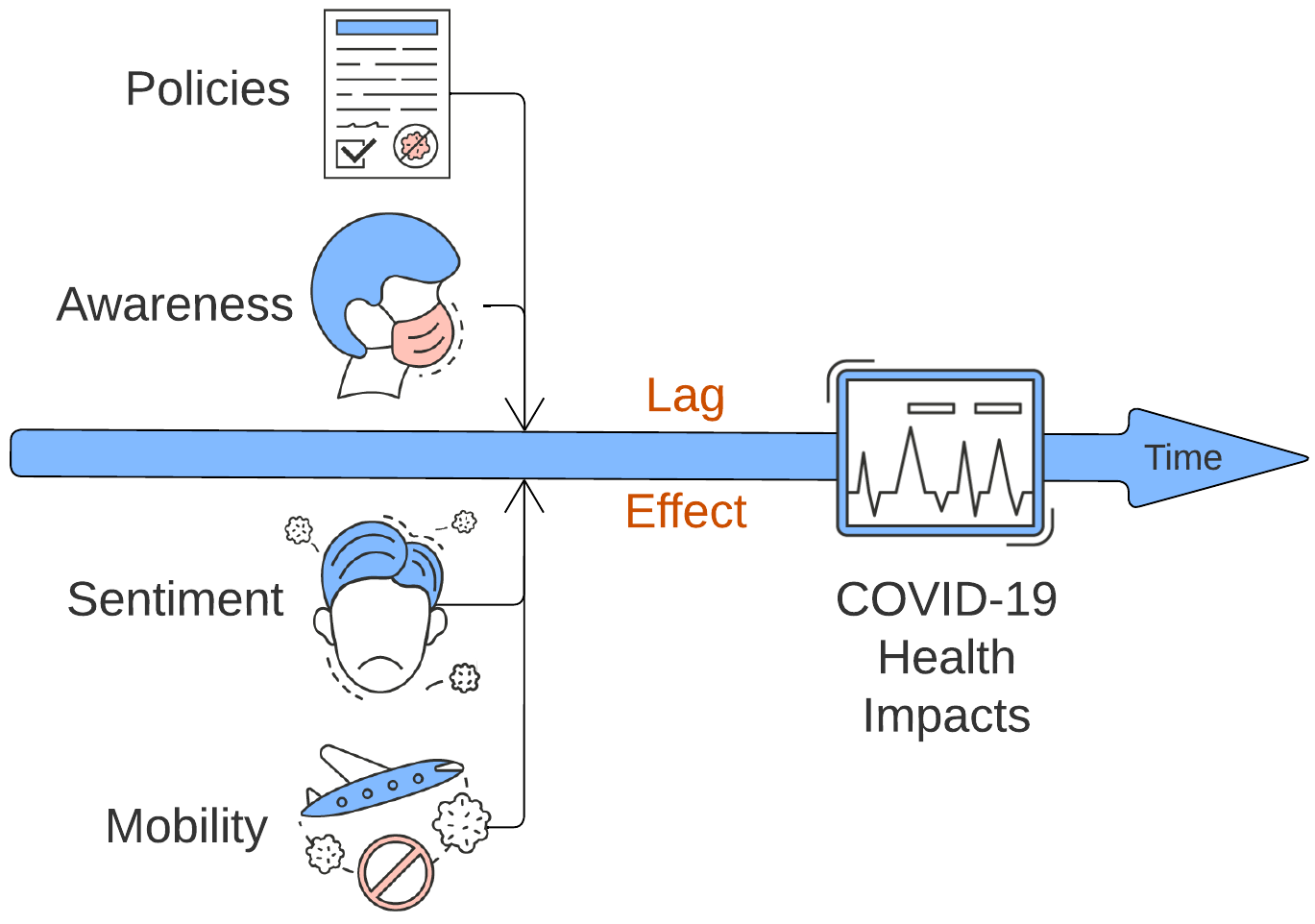}
    \caption{The conceptual framework describing the hypothesized effects of NPIs on COVID-19 health impacts with a time lag effect.}
    \label{figure:Conceptual_Framework}
\end{figure}

Figure \ref{figure:Conceptual_Framework} illustrates the conceptual framework describing the hypothesized effects of NPIs on COVID-19 health impacts with a time lag effect. The stay-at-home policies, public awareness and sentiment toward COVID-19, and human mobility were selected as NPIs in this study. Section 3 outlines the data collection and processing methods employed to measure NPIs and COVID-19 health impacts. 

\subsection{COVID-19 Risk Perceptions }
Twitter, as one of the most popular social media, provides users with a platform to share their experiences, feelings, and opinions about events through short messages (tweets) \citep{zou2019social}. In 2023, Twitter was renamed and branded as X, and the remainder of this article uses Twitter to avoid confusion. With 450 million monthly active users as of 2023 \citep{Twitter450users}, Twitter data become invaluable resources for researchers to quantitatively monitor human perceptions and behaviors during COVID-19 in the near-real time \citep{bogdanowicz2022dynamic, haupt2021characterizing}. However, it is worth mentioning that Twitter data, like many other social media platforms, are inherently biased towards younger, well-educated, and wealthier urban populations \citep{blank2017digital}. Analyzing Twitter data without considering demographic biases might overlook the behavior of certain social groups and lead to unfair estimations.

\begin{figure}[!thbp]
   \centering
   \includegraphics[width=1\textwidth]{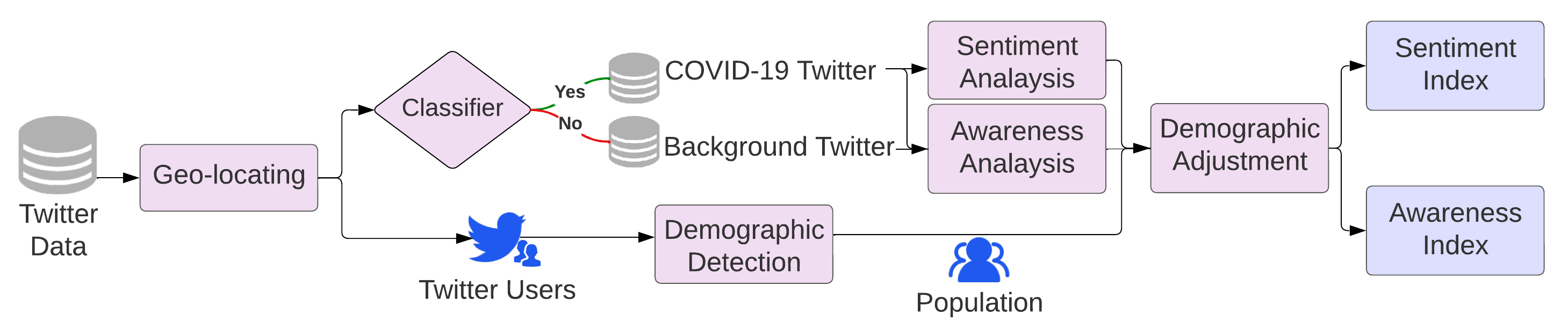}
   \caption{Framework of Twitter data mining for demographically-unbiased assessments of public awareness and sentiment toward COVID-19.}
   \label{fig:twitter}
\end{figure}

To track demographically unbiased public awareness and sentiment toward COVID-19, we conducted the Twitter data mining framework, as depicted in Figure \ref{fig:twitter}. This framework consists of five main steps. First, we collected all geotagged tweets from the U.S. in 2020 using the Twitter Academic Application Programming Interface (API). Non-human generated tweets and tweets from organizational accounts, which were irrelevant to public perceptions, were removed by methods delineated in \cite{lin2022revealing}. A total of 255,291,871 geotagged tweets were retrieved. Second, we set a list of COVID-19-related keywords based on existing literature \citep{banda2021large,alqurashi2020large}, i.e., \textit{covid}, \textit{virus}, \textit{2019-ncov}, \textit{sars-cov-2}, \textit{coronavirus}, \textit{ncov}, \textit{n95}, \textit{social distancing}, \textit{lockdown}, \textit{quarantine}, \textit{pandemic}, \textit{epidemic}, \textit{pneumonia}, and \textit{confirmed cases}, to identify tweets relevant to the pandemic. A total of 3,954,468 tweets (1.55\%) were identified as COVID-19-related. Third, we calculated the percentage of COVID-19-related Twitter data over all geotagged tweets as the Ratio index to represent public awareness toward COVID-19 \citep{lin2022revealing}. In terms of public sentiment toward COVID-19, we estimated users’ sentiment toward COVID-19 (negative, neutral, or positive) based on the sentiment of all COVID-19-related tweets they posted \citep{lin2023sensing}. Fourth, we employed the M3 (multimodal, multilingual, and multi-attribute) model proposed by \cite{wang2019demographic} to detect the demographics of users including age and gender based on users’ screen names, usernames, profile images, and biographies. Finally, the positive and negative Sentiments Adjusted by Demographics (P-SAD and N-SAD) index and Ratio Adjusted by Demographics (RAD) index were computed using the post-stratification method based on the difference between the demographic structure of Twitter users and the general population, as suggested in \citep{lin2023sensing}. The N-SAD and P-SAD indexes represent the demographically unbiased percentages of Twitter users expressing overall negative and positive emotions toward COVID-19, respectively. The RAD index quantifies the proportion of tweets concerning COVID-19 among all tweets after correcting Twitter users' demographic biases.

\subsection{Mobility}
 In this study, we collected daily Apple mobility data in the U.S. at the state level in 2020 to assess human mobility in different modes, namely driving, walking, and public transit. The Apple human mobility data track the mobility volume change in driving, walking, and taking public transit at multiple administrative levels, e.g., global, country, state, and county \citep{Applemobilitydata}. The data were derived from Apple Maps users and reported as the relative volume based on the baseline volume, which was the direction requests received per country/region, sub-region, or city on January 13th, 2020. With the Apple human mobility data, it becomes feasible to monitor spatiotemporal changes in human mobility during significant events like COVID-19. It is worth noting that Apple is no longer offering mobility trends reports as of April 2022.  

\subsection{COVID-19 Policies}
The Oxford COVID-19 Government Response Tracker (OxCGRT), developed by researchers at the University of Oxford using the scorecard method, offers a systematic estimation of the stringency of COVID-19 policies implemented by various countries since January 1st, 2020. \cite{hale2021global} compiled a comprehensive set of policies and assigned scores to each policy, with higher scores indicating more stringent measures. These policies were categorized into 23 indicators based on their thematic focus. The Stringency Index (SI) selected in this investigation quantifies the strictness of stay-at-home COVID-19 policies by incorporating nine indicators, namely school closures, workplace closures, restrictions on public events, limitations on gathering size, public transport closures, stay-at-home requirements, restrictions on internal movement, restrictions on international travel, and public information campaigns. The SI scale ranges from 0 to 100, with higher values indicating more stringent measures. For this study, we collected daily SI data at the state level in the U.S. throughout 2020.

\subsection{COVID-19 Cases}
To assess the health implications of COVID-19, we utilized the case rate as a quantitative measure, which represents the proportion of confirmed cases per 100,000 individuals within the population. The cumulative confirmed cases in the U.S. in 2020 were collected from the publicly accessible database maintained by the Center for Systems Science and Engineering (CSSE) at Johns Hopkins University \citep{dong2020interactive}. Population data were sourced from the United States Census, and estimates were based on data as of April 1st, 2020. The resulting case rate values ranged from 0 to 105, with higher values indicating a more pronounced impact on public health attributed to COVID-19.

\section{Methods}

\subsection{The MBSTS Model}

The MBSTS model is a general time series model constructed as the sum of trend $\mu{(t)}$, season $\tau{(t)}$, cycle $\omega{(t)}$, and regression $\xi{(t)}$ components with $t\in \{1,\ldots, T\}$ being the time index, as follows:
$$
Y{(t)}=\mu{(t)}+\tau{(t)}+\omega{(t)}+\xi{(t)}+\epsilon{(t)}, \quad\quad\quad \epsilon{(t)} \widesim{\text{i.i.d.}} \mathcal{N}_M\left(0, \Sigma_\epsilon\right),
$$
where $Y{(t)}=\{Y_m(t)\}_{m = 1}^{M}$ is an $M$-dimension outcome vectors. All components are assembled independently, with each component yielding an additive contribution. The MBSTS model allows each $Y_m(t)$ for $m\in \{1,\ldots, M\}$ to have its specific formula. For instance, for predicting two-dimension outcome vectors, the first time series may encompass the trend, season, and regression components, while the second time series may only have the trend component. The model training is conducted over all $M$ time series incorporating the correlations through $M\times M$-dimensional covariance associated with the error term $\epsilon{(t)}$. 

The specification of the trend component in a time series model depends on both the characteristics displayed by the analyzed series and any available prior knowledge. If the series consistently demonstrates either an upward or downward movement, incorporating a slope or drift into the trend model could be suitable. This results in a more comprehensive model compared to the local linear trend model. In this generalized version, the slope remains stationary rather than random, and the model can be expressed in the following form:
$$
\begin{aligned}
	\mu(t+1) & =\mu{(t)}+\tilde{\delta}{(t)}+\tilde{u}{(t)}, & \tilde{u}{(t)} \widesim{\text{i.i.d.}} \mathcal{N}_M\left(0, \Sigma_\mu\right), \\
	\tilde{\delta}(t+1) & =\tilde{D}+\tilde{\rho}\left(\tilde{\delta}{(t)}-\tilde{D}\right)+\tilde{v}{(t)}, \hspace{1cm}& \tilde{v}{(t)} \widesim{\text{i.i.d.}} \mathcal{N}_M\left(0, \Sigma_\delta\right).
\end{aligned}
$$
Here, $\tilde{\delta}{(t)}$ and $\tilde{D}$ represent $m$-dimensional vectors. Specifically, $\tilde{\delta}{(t)}$ signifies the expected increase in $\mu{(t)}$ between time $t$ and $t+1$ to resemble a short-term slope at time $t$. In contrast, $\tilde{D}$ pertains to the long-term slope. This structural setup harmonizes short-term insights with long-term trends, resulting in a model that appropriately blends both types of information.

The second component of the model is responsible for capturing seasonality. A commonly utilized model is expressed as follows:
$$
\tau_m(t+1)=-\sum_{k=0}^{S_m-2} \tau_m(t-k)+w_m{(t)}, \quad\quad\quad \tilde{w}{(t)}=\left[w_{1}(t), \cdots, w_{M}(t)\right]^T\widesim{\text{i.i.d.}} \mathcal{N}_M\left(0, \Sigma_\tau\right).
$$
Here, $S_m$ represents the number of seasons for the time-series $Y_m(t)$ for $m\in \{1,\ldots, M\}$, and the $M$-dimensional vector $\tau{(t)}=(\tau_1(t),\ldots, \tau_M(t))$ signifies their collective influence on the observed target time series $Y{(t)}=(Y_1(t),\ldots, Y_M(t))$. The MBSTS model accommodates diverse seasonal components with distinct periods for each target series $Y_m(t)$. For instance, it's possible to incorporate a seasonal component with $S_m=7$ to capture the day-of-the-week effect for one target series, and $S_{m'}=30$ to account for the day-of-the-month effect in another target series.

The third component of the series aims to capture cyclical effects. In economics, the term "business cycle" refers to recurrent deviations around the long-term trajectory of the series that are not strictly periodic. A model encompassing a cyclical component can effectively replicate crucial features of the business cycle, such as robust autocorrelation, alternating phases, damping fluctuations, and null long-term persistence. A stochastic trend model, when applied to a seasonally adjusted economic time series, may not adequately capture the series' short-term fluctuations on its own. However, by integrating a serially correlated stationary component, the model becomes equipped to account for these short-term movements, thereby encompassing the cyclical influence. The cycle component is defined as follows:
$$
\begin{aligned}
	& \omega(t+1)=\varrho \widehat{\cos (\lambda)} \omega{(t)}+\varrho \widehat{\sin (\lambda)} \omega^{\star}{(t)}+\tilde{\kappa}{(t)}, \quad &\tilde{\kappa}{(t)} \widesim{\text{i.i.d.}} \mathcal{N}_M\left(0, \Sigma_\omega\right), \\
	& \omega^{\star}(t+1)=-\varrho \widehat{\sin (\lambda)} \omega{(t)}+\varrho \widehat{\cos (\lambda)} \omega^{\star}{(t)}+\tilde{\kappa}^{\star}{(t)}, \quad\quad &\tilde{\kappa}^{\star}{(t)} \widesim{\text{i.i.d.}} \mathcal{N}_M\left(0, \Sigma_\omega\right), 
\end{aligned}
$$
where $\varrho, \widehat{\sin (\lambda)}$, and $\widehat{\cos (\lambda)}$ are $M \times M$ diagonal matrices with diagonal entries equal to $\varrho_{i i}$ (a damping factor for target series $Y_{i}$ such that $\left.0<\varrho_{i i}<1\right), \sin \left(\lambda_{i i}\right)$ where $\lambda_{i i}=2 \pi / q_i$ is the frequency with $q_i$ being a period such that $0<\lambda_{i i}<\pi$, and $\cos \left(\lambda_{i i}\right)$ respectively. 

The regression component $\xi{(t)}=(\xi_{1}{(t)},\ldots,\xi_{M}{(t)})$ with static coefficients is written as follows:
\begin{equation}
	\xi_{m}{(t)}=\beta_m^T X_{m}{(t)} .
	\label{eq:intepret}
\end{equation}
Here, $\xi_{m}{(t)}=\left[\xi_{m,1}{(t)}, \cdots, \xi_{m,d}{(t)}\right]^T$ is the collection of all elements in the regression component. For target series $Y_{m}$, $X_{m}{(t)}=\left[X_{m,1}{(t)}, \ldots, X_{m,d}{(t)}\right]^T$ is the pool of predictors at time $t$, and $\beta_m=\left[\beta_{m,1}, \ldots, \beta_{m,d}\right]^T$ represents corresponding static regression coefficients. Regression analysis is a statistical methodology used to estimate relationships between dependent variables and independent variables, which are alternatively referred to as predictors, covariates, or features. That is, each time series has its specific $d$ predictors that are different from those of other time series. The total number of different predictors for the $M$-dimensional time series is thus $Md$. 

The MBSTS model is able to select important features while taking into account the spatial correlations among target time series, by means of the spike and slab technique developed by \cite{george1997approaches} and \cite{madigan1994model} that has been widely used for dimension reduction \citep{jammalamadaka2019predicting, qiu2020multivariate, ning2023bayesian}.
Additionally, the model is equipped to infer the trend component $\mu{(t)}$, the seasonal component $\tau{(t)}$, and the cyclical component $\omega{(t)}$ via a posterior simulation algorithm as outlined by \cite{durbin2002simple}. Moreover, it enables the inference of covariance matrices associated with these three components, namely, $(\Sigma_\mu,\Sigma_\delta,\Sigma_\tau,\Sigma_\omega)$ through an inverse Wishart distribution.

\subsection{The MBSTS-TL algorithm} \label{sec:algorithm}
Although the MBSTS model is suitably employed to detect the compounding effects of NPIs on pandemic transmission, it is imperative to recognize that the influence of NPIs on the epidemic may exhibit delayed effects. There is a need to modify the MBSTS model, which currently assumes that factors affect the target time series instantaneously, by incorporating time lag effects associated with NPIs. This augmentation is critical for improving the model's capacity to faithfully capture real-world dynamics. Therefore, in this subsection, we propose a new Algorithm \ref{algorithm 1}, named MBSTS-Time Lagged (MBSTS-TL) model, designed to work with the MBSTS model for the spatiotemporal setting with a time lag of $l_t$, within which we introduce a proper error metric for evaluation and hyper-parameter tuning. Effective hyper-parameter selection is crucial in spatiotemporal analysis. Notably, this work represents the first attempt to provide an explicit hyper-parameter tuning method within the MBSTS framework.

The novel error metric, denoted as $\operatorname{AE}_{\rho, \text{S}, \varrho, \lambda}$, facilitates the tuning of hyper-parameters ($\rho$, $S$, $\varrho$, and $\lambda$) associated with the MBSTS model. It takes into account both temporal variations and spatial disparities. This methodology identified a diverse set of candidate parameters and selected the optimal one to enhance our model, which is straightforward to implement. 

To elaborate, we divide the time interval $[t^{\text{start}}, t^{\text{end}}]$ into non-overlapping $K$ segments, creating a partition as follows:
$$[t_1^{\text{start}},t_1^{\text{end}}],\;[t_2^{\text{start}},t_2^{\text{end}}],\; \ldots\;, [t_K^{\text{start}},t_K^{\text{end}}]$$
This partition allows us to evaluate model performance in distinct time stages, considering the error metric defined in Equation \eqref{eq:absolute error}. Importantly, this time-based partitioning does not increase the computational complexity in the MBSTS model. Given that MCMC, an offline method, is employed, training the MBSTS model with all time-dependent data might lead to lengthy convergence times and require substantial computational resources. Our approach addresses this issue by allowing users to define time partitions that align with the evolving dynamics of events, such as the varying stages of COVID-19 spread. For each MBSTS model corresponding to a partition segment, a smaller-scale MCMC is performed. This not only makes the process computationally feasible on personal computers but also adapts the model to the evolving nature of spatiotemporal phenomena.

\begin{algorithm}[!ht] \caption{The MBSTS-TL algorithm} \label{algorithm 1} 

    \begin{algorithmic}

    \State \textbf{INPUT:} Covariate $X(t^{\text{start}}_k), \ldots, X(t^{\text{end}}_k - l_{t})$ and outcome $Y(t^{\text{start}}_k+ l_{t}), \ldots, Y(t^{\text{end}}_k)$ for
    
     \hspace{1cm}$k \in \{ 1, ... ,K\}$.
    \smallskip
    
    \State \textbf{Evaluation:} 
    
    \begin{enumerate}
    \item Training the $k$-th MBSTS model for $k = 1, ... ,K$, with hyper-parameter $\rho$, $S$, $\varrho$, and $\lambda$ using
    $$X(t^{\text{start}}_k), \ldots, X(t^{\text{end}}_k - 1 - l_{t})\quad \text{and} \quad Y(t^{\text{start}}_k+ l_{t}), \ldots, Y(t^{\text{end}}_k - 1).$$

    \item One step prediction of $Y(t^{\text{end}}_k)$ with $X(t^{\text{end}}_k - l_{t})$ using the trained $k$-th MBSTS model with hyper-parameter $\rho$, $S$, $\varrho$, and $\lambda$. Denote the prediction as $\widehat{Y}(t^{\text{end}}_k)$, for $k = 1, ... ,K$.

   \item Compute the normalized absolute values of the differences between the true values $Y(t^{\text{end}}_k)=\{Y_m(t^{\text{end}}_k)\}_{m = 1}^{M}$ and its corresponding predicted values $\widehat{Y}(t^{\text{end}}_k)=\{\widehat{Y}_m(t^{\text{end}}_k)\}_{m = 1}^{M}$, i.e., 
\begin{equation}
	\operatorname{AE}_{\rho, \text{S}, \varrho, \lambda}(l_t) = \left(\frac{\sum_{m = 1}^{M}|{\widehat{Y}_{m}(t^{\text{end}}_1)}- {Y_{m}(t^{\text{end}}_1)}|}{{M\max_{m\in \{1,\ldots,M\}}{{Y_{m}(t^{\text{end}}_1)}}}}, \,\ldots, \,\frac{\sum_{m = 1}^{M}|{\widehat{Y}_{m}(t^{\text{end}}_K)}- {Y_{m}(t^{\text{end}}_K)}|}{{M\max_{m\in \{1,\ldots,M\}}{{Y_{m}(t^{\text{end}}_K)}}}}\right). \label{eq:absolute error}
\end{equation}
 \end{enumerate}    
 \smallskip
 
		\State \textbf{Training:}
		
		\begin{enumerate}
			\item  Grid search for the optimal hyper-parameters $\rho^*$, $S^*$, $\varrho^*$, and $\lambda^*$ in their user-defined spaces that yield the minimum $\operatorname{AE}_{\rho, \text{S}, \varrho, \lambda}(l_t)$ for different $l_t$.
			\item Generate regression coefficients ${\beta_k}=\left[\beta_{k, 1}, \ldots, \beta_{k, d}\right]^{\top}$ and its confidence interval (CI) for $k = 1, \cdots, K$.
		\end{enumerate}
 \smallskip
 		
		\State \textbf{OUTPUT:} parameters ${\beta_k}$, its CI, and predictions $\widehat{Y}^*(t^{\text{end}}_k)$ for $k = 1, ... ,K$, and error 
		
		\hspace{1.4cm}$\operatorname{AE}_{\rho^*, \text{S}^*, \varrho^*, \lambda^*}(l_t)$.
		
	\end{algorithmic}
	
\end{algorithm}

The time interval of $53$ weeks during the year 2020 was divided into three segments: $[9, 22]$, $[23, 37]$, and $[38, 53]$, representing February 24$^{th}$ to May 31$^{st}$, June 1$^{st}$ to September 13$^{th}$, and September 14$^{th}$ to December 31$^{st}$, 2020. These periods, corresponding to the onset of the outbreak, the phase of rapid spread, and the full-blown phase of the pandemic, were named as the initial outbreak, rapid spread, and full-blown periods, respectively.

\section{Results}

\subsection{Temporal trends of COVID-19 health impacts and NPIs in the U.S.}
\begin{figure}[H]
    \centering
    \includegraphics[width=1\textwidth]{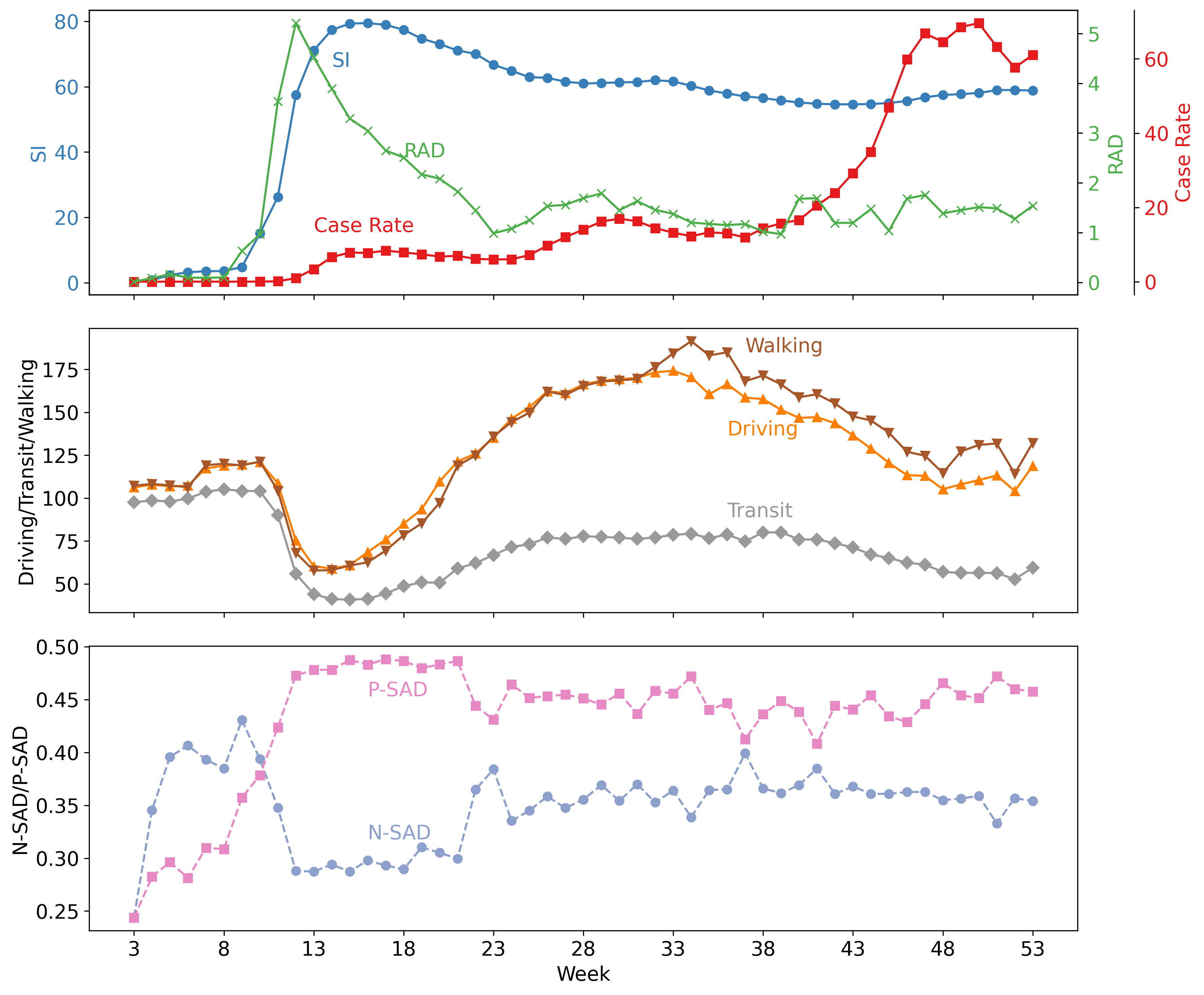}
    \caption{National temporal trends of COVID-19 health impacts and NPIs}
    \label{fig:timeseries}
\end{figure}

Figure \ref{fig:timeseries}  illustrates the national temporal trends of COVID-19 health impacts and NPIs from week 3 to week 53 (mid-January to the end of December) in 2020 within the United States.  The case rate exhibited a fluctuation spanning from 0 to 69.62, characterized by three distinct stepwise increments. There were two minor ascensions of approximately 10 each, occurring between weeks 13 to 22 (late March to the end of May) and weeks 23 to 37 (June to mid-September), as well as a rapid ascent to approximately 60 between weeks 38 to 47 (mid-September to late November). Subsequently, the case rate dynamically sustained itself around 60 between weeks 48 and 53 (late November to the end of December). The RAD index ranged from 0.01 to 5.21. It remained proximate to 0 before week 8 (mid-February), after which it underwent a rapid ascent commencing in week 11 (mid-March), reaching its zenith in week 12, followed by a descent to 0.99 by week 23 (early June). Thereafter, it maintained values around 1.5. The SI index ranged from 0.31 to 79.45. Prior to week 9 (the end of February), values remained below 5, but subsequent to week 10 (early March), a rapid ascent was observed, reaching its pinnacle at week 16 (mid-April), followed by a gradual descent. Values around 60 were sustained from week 25 (mid-June) to week 53 (the end of December).

Regarding the variations in human mobility, the walking Index ranged from 57.74 to 191.47, the driving Index ranged from 58.88 to 174.19, and the transit Index ranged from 40.75 to 80.04. These three indices exhibited similar trends, with each maintaining relatively stable values from weeks 3 to 10 (mid-January to early March), at 110, 110, and 100, respectively. A decline was initiated from week 11 (mid-March), reaching respective minima in weeks 13 (the end of March), 14, and 15 (early April), followed by an ascent to their peaks from weeks 28 to 38 (early July to mid-September). Finally, they gradually decreased to approximately 125, 110, and 57 by weeks 48 to 53 (late November to the end of December). In general, the trends for walking and driving exhibited a high degree of overlap, characterized by an early rapid decline, recovery, and surpassing of the normal baseline values. The usage of public transits exhibited a more pronounced initial decline compared to walking and driving, with subsequent recovery, and it did not return to the values observed in the normal status, maintaining an overall lower value.

For the characterization of sentiment, the N-SAD index ranged from 0.24 to 0.43. Prior to week 9 (late March), it exhibited an upward trajectory, increasing from a minimum of 0.24 to a maximum of 0.43. Thereafter, a rapid decline ensued, reaching 0.29 by week 12 (mid-March). From weeks 12 to 21 (mid-March to late May), the index remained at around 0.3, subsequently stabilizing at approximately 0.35 from week 22 to week 53 (late May to the end of December). The P-SAD index ranged from 0.24 to 0.49. It underwent a rapid ascent from week 3 to week 12 (mid-January to mid-March), increasing from 0.24 to 0.47, and subsequently maintained values around 0.47 from week 13 to week 21 (late March to late May), followed by values around 0.45 from week 22 to week 53 (late May to the end of December). Notably, during the initial 3 to 10 weeks (mid-January to early March), corresponding to the period prior to the COVID-19 outbreak in the United States, the proportion of Twitter users expressing negative sentiment towards the pandemic was higher than those with a positive sentiment. Conversely, after the outbreak, the proportion of users expressing negative sentiment rapidly declined, while those with  positive sentiment exhibited a substantial increase. After week 11 (mid-March), the proportion of Twitter users with a positive sentiment consistently exceeded those with a negative sentiment.

\subsection{Spatiotemporal disparities of COVID-19 health impacts and NPIs}
\begin{figure}[!thbp]
    \centering
    \includegraphics[width=1\textwidth]{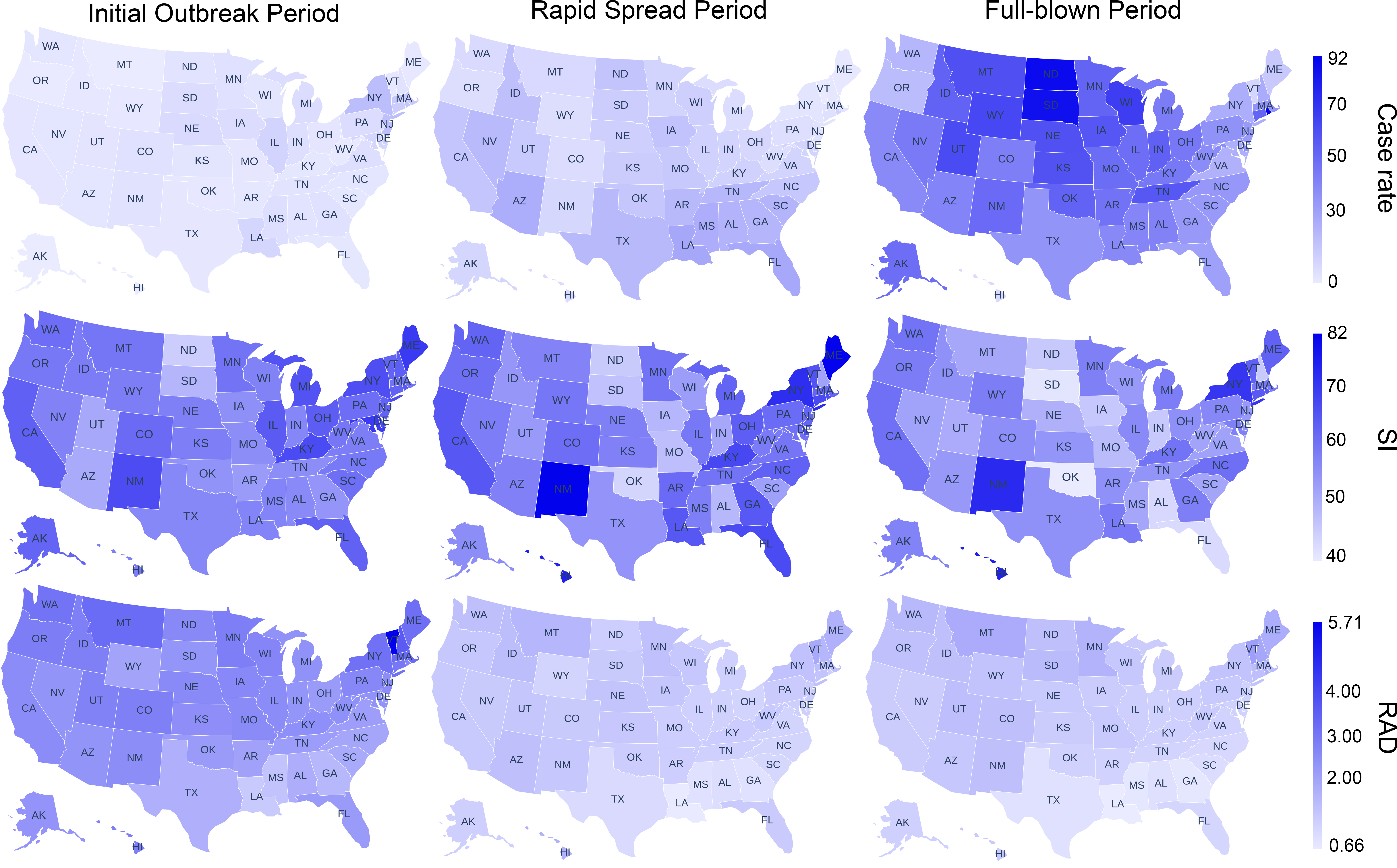}
    \caption{Spatiotemporal disparities of COVID-19 health impacts, SI and RAD}
    \label{fig:datafigure_caserate_SI_Ratio}
\end{figure}

Figure \ref{fig:datafigure_caserate_SI_Ratio} depicts the spatial disparities of COVID-19 case rates, SI index, and RAD index at the state level across three distinct phases in 2020: the initial outbreak period, the rapid spread period, and the full-blown period.

During the initial outbreak period, the case rate ranged from 0 to 19, with New York state exhibiting the highest case rate. In the rapid spread period, case rates ranged from 1.06 in Vermont to 27.54 in Louisiana. States in the southern region of the United States, such as Florida (26.88), Arizona (25.11), and Mississippi (24.87), demonstrated notably higher case rates. In the full-blown period, case rates spanned from 8.68 in Vermont to 89.45 in North Dakota, with northern states like South Dakota (86.98), Wisconsin (68.90), and Wyoming (65.23) displaying higher case rates.

The SI index during the initial outbreak period ranged from 45.41 in North Dakota to 71.92 in Maine. States with relatively higher SI indices included Maryland (71.71), Kentucky (68.85), Delaware (68.81), New Mexico (68.66), and New York (68.27). Conversely, states with lower SI indices encompassed South Dakota (49.79), Arizona (51.75), and Utah (52.04). In the rapid spread period, the SI index varied from 43.98 in Oklahoma to 82.05 in New Mexico, with states like Maine (81.95), Hawaii (76.85), New York (74.03), and Kentucky (69.42) displaying higher SI indices. Lower SI indices were observed in states like North Dakota (46.04), South Dakota (47.62), and Missouri (48.14). In the full-blown period, the SI index ranged from 40.02 in Oklahoma to 76.61 in Hawaii. States with higher SI indices included New Mexico (74.85), New York (72.66), and Connecticut (66.27), while states with lower SI indices were South Dakota (41.31), Florida (43.03), and Alabama (43.52). Overall, New Mexico and New York maintained higher SI indices throughout the three phases, suggesting stricter stay-at-home policies in these states. North Dakota and South Dakota consistently exhibited lower SI indices across the three phases, indicating more relaxed stay-at-home policies.

The RAD index during the initial period ranged from 1.39 in Louisiana to 5.71 in Vermont, with other states such as New Hampshire (3.78) and Massachusetts (3.46) exhibiting higher RAD indices. Lower RAD indices were observed in states like Louisiana (1.39) and Mississippi (1.53). In the rapid spread period, the RAD index ranged from 0.69 in Louisiana to 2.10 in Vermont, with higher RAD indices in Hawaii (1.91) and Maine (1.91). Lower RAD indices were observed in Georgia (0.87) and Mississippi (0.88). In the full-blown period, the RAD index ranged from 0.66 in Louisiana to 2.31 in Vermont, with higher RAD indices in New Hampshire (2.13) and Montana (2.04). Lower RAD indices were observed in Georgia (0.76) and Mississippi (0.76). It is evident that Louisiana consistently exhibited lower RAD indices across all three phases, indicating lower levels of public awareness of the pandemic among Twitter users in that state.

\begin{figure}[H]
    \centering
    \includegraphics[width=1\textwidth]{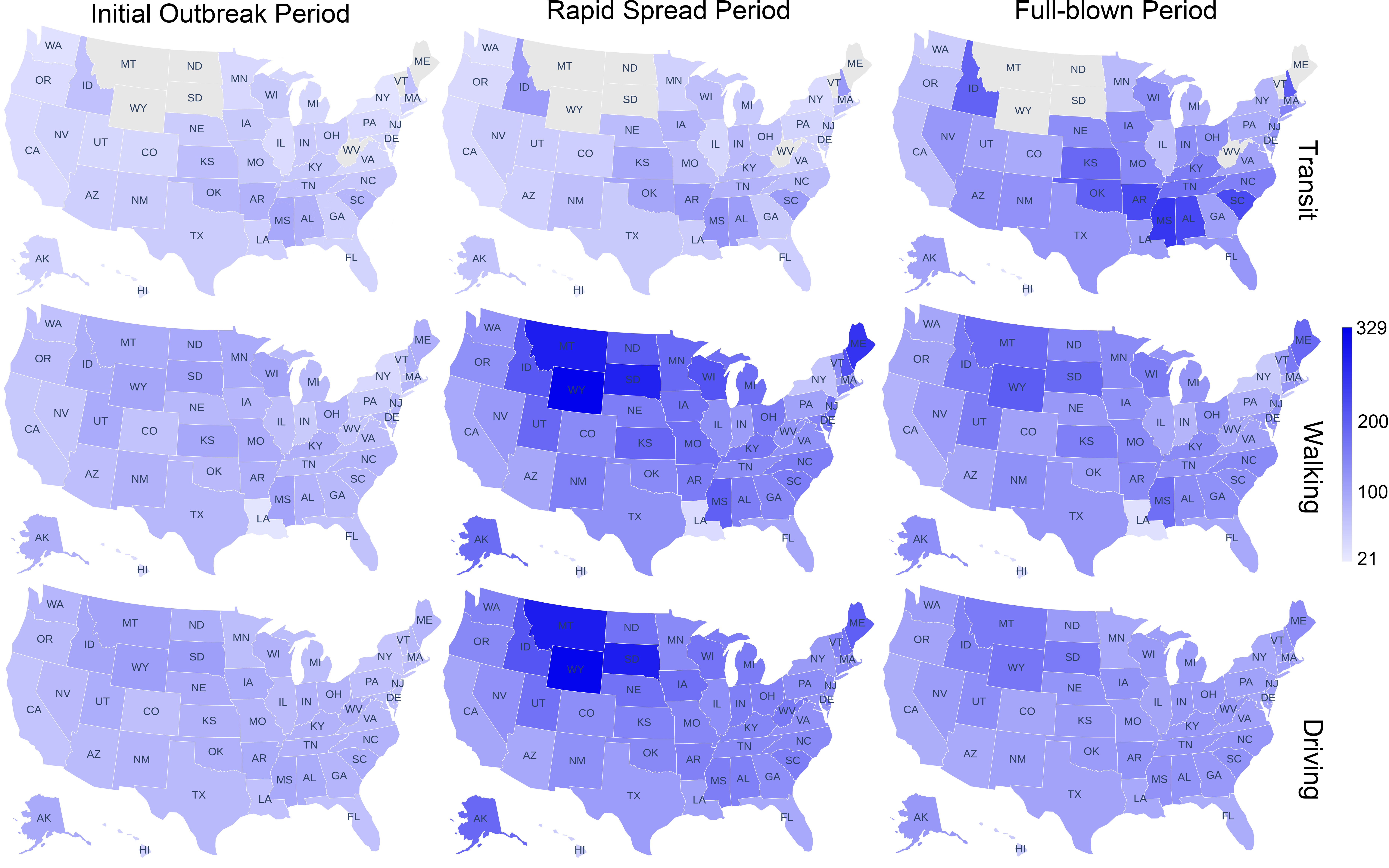}
    \caption{Spatiotemporal disparities of human mobility}
    \label{fig:datafigure_mobility}
\end{figure}

Figure \ref{fig:datafigure_mobility}  illustrates the spatial disparities in human mobility categorized into three modes: Transit, Walking, and Driving at the state level during three distinct periods of the pandemic. Areas depicted in gray signify the unavailability of data. During the initial outbreak period, Transit ranged from 30.57 (Hawaii) to 107.85 (Mississippi). In addition to Hawaii, New York and Washington displayed lower Transit values of 32.43 and 34.51, respectively. In contrast, Alabama and Arkansas displayed higher Transit values of 97.18 and 95.93, respectively, along with Mississippi. Notably, Mississippi was the sole state with a Transit value exceeding 100, signifying a general reduction in Transit mobility across states due to the pandemic's impact. In the rapid spread period, Hawaii continued to exhibit the lowest Transit value (21.55), while Mississippi maintained the highest value (135.68). New York and Washington still maintained lower Transit values of 39.19 and 42.87, respectively. This indicates that the impact of the pandemic on Transit mobility persisted longer in Hawaii, New York, and Washington. During the full-blown period, Transit ranged from 26.63 (Hawaii) to 109.97 (Mississippi). Alabama (101.86) and Arkansas (100.62) reported Transit values exceeding 100, and New Hampshire and South Carolina also recorded values surpassing 90. This suggests that only five states had Transit mobility nearly returned to pre-pandemic normalcy.

As for Walking and Driving mobility, they exhibited a similar spatiotemporal pattern. During the initial outbreak period, Driving ranged from 55.70 (Hawaii) to 121.68 (South Dakota), while Walking ranged from 30.79 (Louisiana) to 116.60 (South Dakota). In the rapid spread period, Driving ranged from 60.67 (Hawaii) to 324.86 (Wyoming), with all states, except Hawaii, reporting Driving values exceeding 100. Walking ranged from 43.25 (Louisiana) to 328.53 (Wyoming), and only Hawaii and New York reported values below 100, at 43.65 and 71.77, respectively. The remaining states exhibited values exceeding 100, with Wyoming, Montana, South Dakota, and Maine displaying notably high values. In the full-blown period, Walking ranged from 53.00 (Hawaii) to 212.64 (Wyoming), while Driving ranged from 64.63 (Hawaii) to 175.21 (Wyoming). Wyoming, Montana, South Dakota, and Maine continued to exhibit high Walking and Driving values.

Hawaii is the only state where Transit, Walking, and Driving values remained consistently below 100 in all three phases. This signifies that residents of Hawaii experienced the most significant impact on their human mobility due to the pandemic, and recovery over a year has proven challenging.

\begin{figure}[!thbp]
    \centering
    \includegraphics[width=1\textwidth]{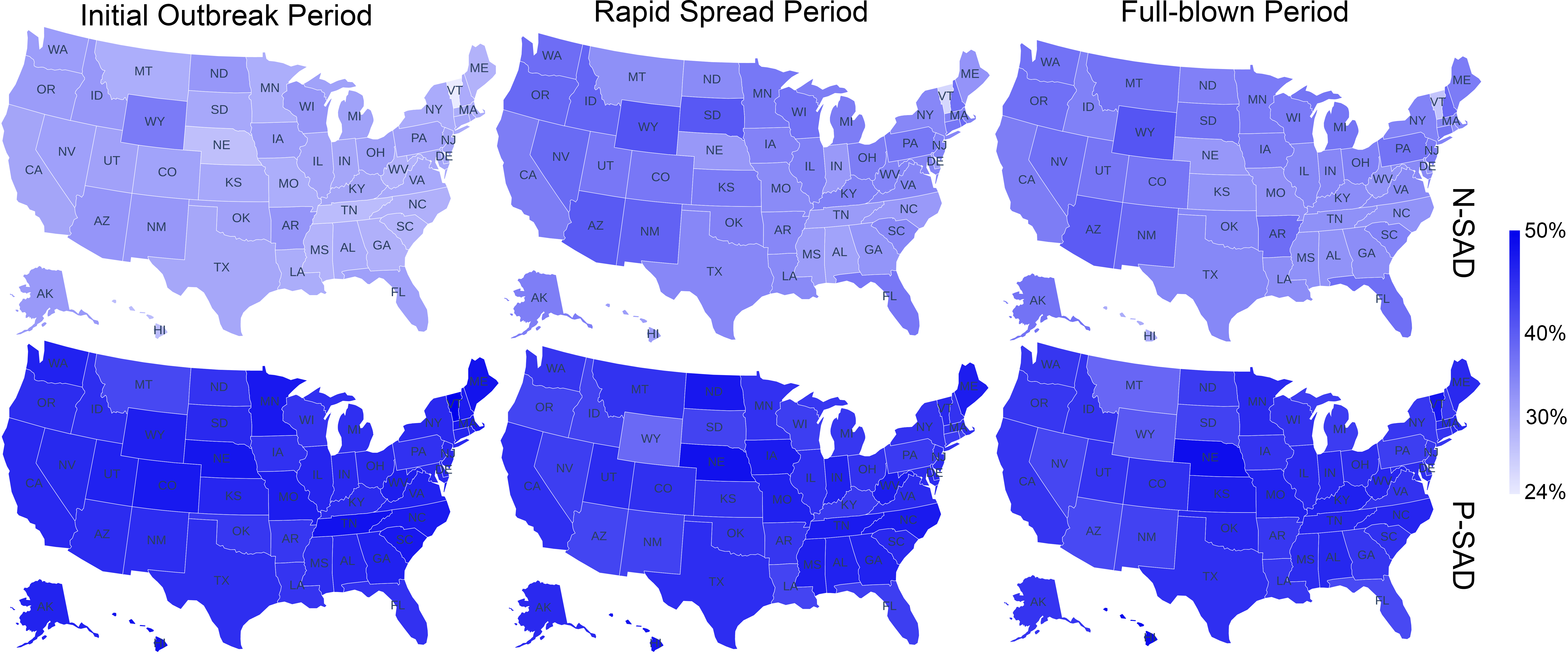}
    \caption{Spatiotemporal disparities of public sentiment toward COVID-19}
    \label{fig:datafigure_sentiment}
\end{figure}

Figure \ref{fig:datafigure_sentiment} depicts the spatiotemporal disparities in public sentiment towards COVID-19 across three distinct periods. The N-SAD index reached its lowest values in Vermont, with values of 0.25, 0.26, and 0.29 during the three periods, respectively. The N-SAD index was highest in Wyoming, with values of 0.37, 0.41, and 0.41 during these periods, respectively. With respect to the P-SAD index, during the initial outbreak period, it displayed a range from 0.42 (Montana) to 0.50 (Vermont). In the rapid spread period, the P-SAD index exhibited variations ranging from 0.39 (Wyoming) to 0.48 (Nebraska), while during the full-blown period, it spanned from 0.39 (Montana) to 0.49 (Hawaii). The Twitter user sentiment within Vermont consistently reflected a more optimistic outlook toward the pandemic, while users in Wyoming generally expressed a more pessimistic sentiment.

\subsection{Model performance and time lag effects}
In this study, we set the feature $X_m$ to be SI, driving, transit, walking, RAD, negative, and positive and set the outcome $Y_m$ to be the COVID-19 case rate, for each state $m\in \{1,\ldots,50\}$ in the United States. 
For hyper-parameter tuning, we used grid search with the following parameter grids: 
$\mathbb{S}_\rho = (0.2, 0.4, 0.6, 0.8)$,  
$\mathbb{S}_S = (3, 4, 5, 6, 8, 10, 12)$, 
$\mathbb{S}_\rho  = (0.1, 0.2, 0.4, 0.6, 0.8, 0.9)$, 
$\mathbb{S}_\lambda  = (0, \frac{\pi}{2}, \pi)$. 
The optimal choices of tuning parameters were obtained by applying Algorithms \ref{algorithm 1}.

Meanwhile, we introduced temporal lag effects into the BSTS model, denoting this enhanced version as the BSTS-TL model. BSTS-TL model serves as the baseline model to assess the efficacy of the newly designed MBSTS-TL model. We subsequently present a comparative evaluation of the modeling performance of both the BSTS-TL and MBSTS-TL models for three periods (the initial outbreak period, the rapid spread period, and the full-blown period), considering no time lag, 1-week lag, and 2 weeks lag effects through the year, i.e. $l_{t}  = (0, 1, 2)$.

Table \ref{table:BSTS} reports the normalized absolute error \eqref{eq:absolute error} generated by the BSTS-TL model across different time lag selections and different pandemic phases. During the initial outbreak period, the normalized absolute errors range from 0.188 to 0.194, while in the rapid spread period, they range from 0.067 to 0.074. In the full-blown period, the normalized absolute errors range from 0.084 to 0.103. The average normalized absolute errors for time lags of 0, 1, and 2 weeks are 0.116, 0.115, and 0.117, respectively.
As indicated in Table \ref{table:lageffect}, the MBSTS-TL model demonstrates much smaller average normalized absolute errors across all three time lag selections. Specifically, the average normalized absolute errors for time lags of 0, 1, and 2 weeks are 0.044, 0.047, and 0.051 respectively. During the initial outbreak period, the normalized absolute errors range from 0.016 to 0.023. In the rapid spread period, they range from 0.017 to 0.019, and in the full-blown period, they range from 0.097 to 0.111. 

Concurrently, the MBSTS-TL model demonstrates high accuracy during the initial outbreak and rapid spread periods, with consistently low normalized absolute errors. However, its performance is relatively less accurate during the full-blown period. This observation aligns with the reality that because NPIs and COVID-19 spread in each region were highly volatile during this phase, the correlation becomes less significant comparatively in modeling. 

\begin{figure}[!thbp]
    \centering
    \includegraphics[width=1\textwidth]{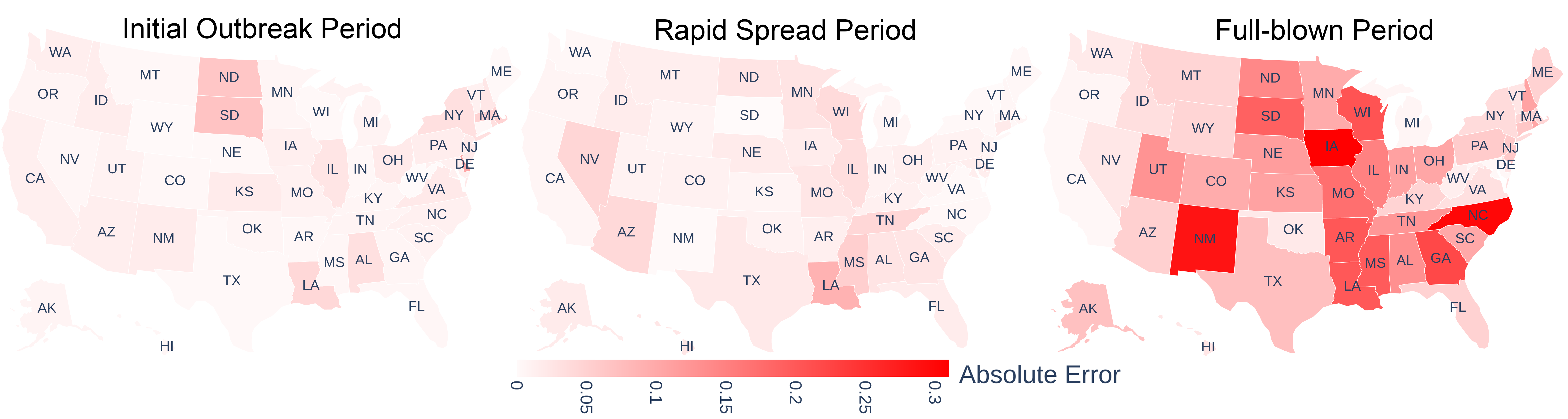}
    \caption{The normalized absolute error of the MBSTS-TL Model at the state level in three pandemic periods}
    \label{fig:PredictionMap}
\end{figure}

Regarding the time lag selection in the MBSTS-TL Model, the smallest normal absolute errors were observed when no time lag was considered during the initial outbreak and full-blown periods, yielding values of 0.016 and 0.097, respectively. In the rapid spread period, the inclusion of a one-week lag led to the smallest normalized absolute error of 0.017 in the MBSTS-TL model. These outcomes indicate the absence of time lag effects of NPIs on the spread of COVID-19 during the initial outbreak and full-blown periods. In contrast, during the rapid spread period, changes in human responses appeared to impact the pandemic's spread one week later. Figure \ref{fig:PredictionMap} portrays the spatiotemporal variation of normalized absolute errors in the MBSTS-TL Model at the state level during the three pandemic periods, considering 0-, 1-, and 2-week time lags. The recorded normalized absolute errors range from 0.00 to 0.31. The highest normalized absolute error emerged in Iowa (IA), North Carolina (NC), and New Mexico (NM) during the full-blown period, yielding respective values of 0.31, 0.30, and 0.29. 
\begin{table}[!thbp]
\caption{The normalized absolute error \eqref{eq:absolute error} in different time lags in the BSTS-TL model}
\centering
\resizebox{\linewidth}{!}{
\begin{tabular}{c|ccc|c|}
\toprule
       Time lag (week) & The initial outbreak period & The rapid spread period & The full-blown period & Average\\ 
    \midrule
        $l_t = 0$ & 0.194 & 0.074 & 0.084 & 0.116\\ 
    \midrule
        $l_t = 1$ & 0.188 & 0.067 & 0.091 & 0.115\\ 
    \midrule
        $l_t = 2$ & 0.194 & 0.067 & 0.091 & 0.117 \\
\bottomrule
\end{tabular}
}
\label{table:BSTS}
\end{table}

\begin{table}[!thbp]
\caption{The normalized absolute error \eqref{eq:absolute error} in different time lags in the MBSTS-TL Model}
\centering
\resizebox{\linewidth}{!}{
\begin{tabular}{c|ccc|c|}
\toprule
      Time lag (week) & The initial outbreak period & The rapid spread period & The full-blown period & Average\\ 
    \midrule
        $l_t = 0$ & 0.016 & 0.018 & 0.097 & 0.044\\ 
    \midrule
        $l_t = 1$ & 0.023 & 0.017 & 0.103 & 0.047\\ 
    \midrule
        $l_t = 2$ & 0.022 & 0.019 & 0.111 & 0.051 \\
\bottomrule
\end{tabular}
}
\label{table:lageffect}
\end{table}

\subsection{Modeled relationships and interpretation}

\begin{figure}[!thbp]
	\centering
	\includegraphics[width=0.91\textwidth]{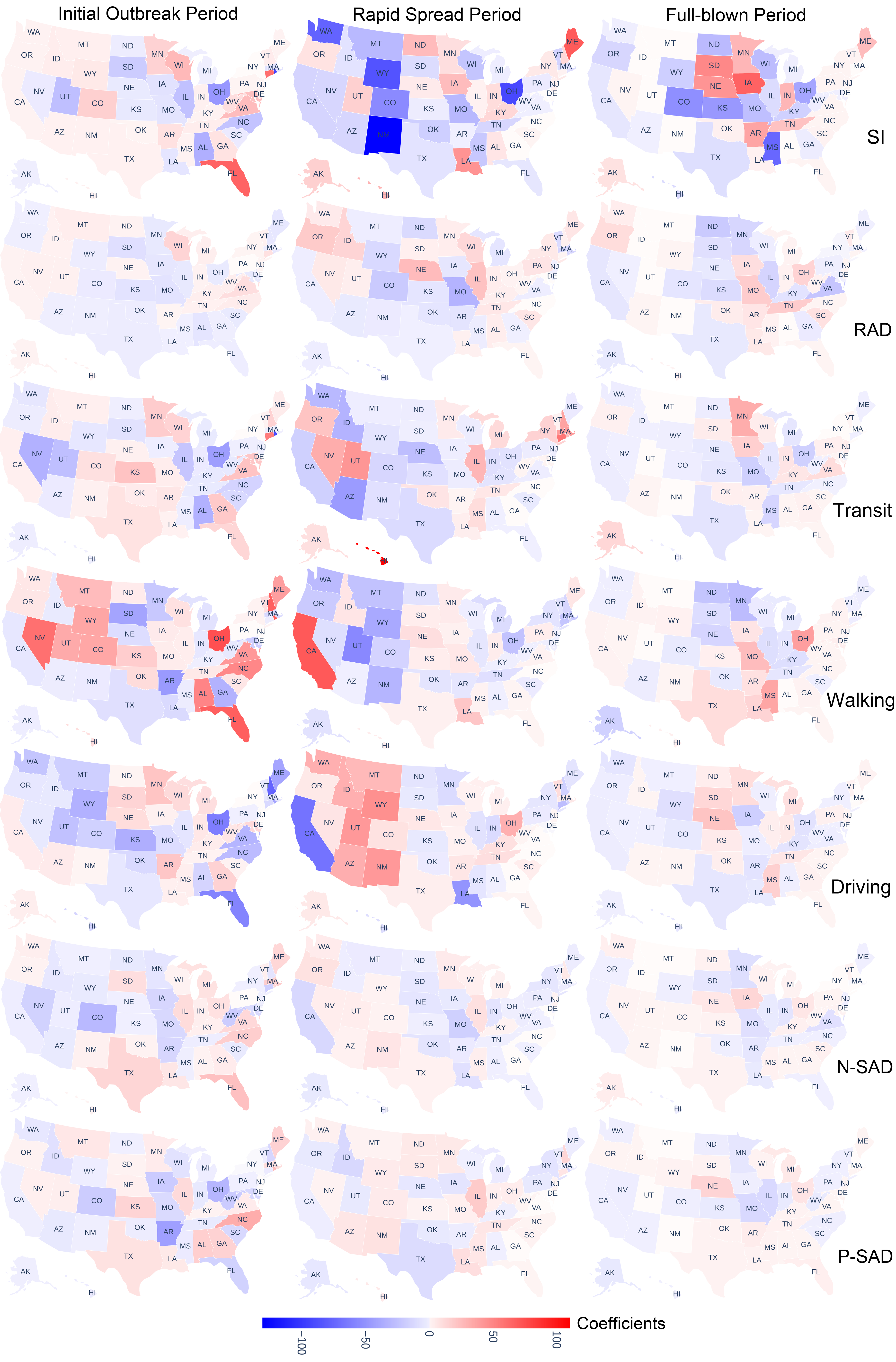}
	\caption{The spatial distribution of coefficients in three phases.}
	\label{fig:CoeffMap}
\end{figure}

Figure \ref{fig:CoeffMap} depicts the coefficients of NPIs derived from the MBSTS-TL model at the state level during three COVID-19 stages. The NPIs coefficients exhibited a range from -130.62 to 109.62, with positive coefficients depicted in red and negative coefficients in blue. These colors signify the positive and negative influences of NPIs on COVID-19 spread, respectively. The saturation of colors reflects the absolute magnitudes of the coefficients, which indicate the degree of significance of NPIs' impacts on the propagation of COVID-19.

Regarding the variations in the primary NPIs influencing the spread of COVID-19 across distinct stages of the pandemic, our analysis identified human mobility in walking and driving as the two key factors impacting case rates in most states during the initial outbreak period. Notably, the coefficients associated with human mobility in walking and driving exhibited the most substantial absolute values across 22 and 13 states, respectively. In particular, the coefficients of human mobility in walking ranged from 74.16 to 33.19 in a decreasing order of the states: Ohio, Florida, Nevada, Alabama, North Carolina, Virginia, Wyoming, Colorado, and Utah. This suggests that in these nine states, an increase in pedestrian movement intensified the COVID-19 outbreak. Conversely, in Arkansas, South Dakota, and Georgia, the coefficients of human mobility in walking ranged from -46.46 to -33.30 in increasing order, indicating that increases in pedestrian mobility alleviated COVID-19 transmission. Furthermore, in states such as New Hampshire, Maine, and Kansas, the coefficients of driving mobility varied from -75.97 to -35.81  in increasing order. These values highlight the association between decreased driving mobility and an uptick in COVID-19 case rates. Overall, the analysis emphasizes the intricate interplay between specific forms of human mobility and the dynamics of COVID-19 spread during different stages of the pandemic.

During the rapid spread period, the coefficients of SI and human mobility by transit displayed the highest absolute values across 20 and 15 states, respectively. Coefficients of SI were negative in five states—New Mexico, Ohio, Wyoming, Washington, and Colorado—ranging from -130.62 to -54.63 in increasing order. Conversely, in Maine, the coefficient of SI was 68.86, unfolding the positive impacts of stay-at-home policy strictness on the case rate change. Regarding human mobility by transiting, the coefficients were negative in Arizona (-45.98) and Idaho (-32.99), and positive in Hawaii, Connecticut, Massachusetts, New Hampshire, and Nevada, ranging from 109.62 to 32.28. It is worth mentioning that in California, human mobility in walking and driving were two principle NPIs with coefficients of 69.52 and -67.78, respectively. These observations reveal that compared to the strictness of stay-at-home policies, the decrease in walking mobility and increment of driving mobility were more forceful NPIs for the inhibition of COVID-19 in California. 

Three NPIs, i.e., stay-at-home policies, human mobility in walking, and public awareness toward COVID-19 served as the key factors of case rate evolving in 18, 10, and 9 states in the full-blown period, respectively. The more strictness of stay-at-home policies coincided with the dwindles of case rate change in Mississippi, Colorado, Kansas, Ohio, and North Dakota with the coefficients of SI ranging from -74.13 to -40.51 in an increasing order. Contrarily, the strictness of stay-at-home policies fostered positive impacts on COVID-19 spread in Iowa, South Dakota, Nebraska, and Arkansas, with the coefficients of SI ranging from 65.68 to 36.29. With respect to walking mobility, the coefficients were 14.61 in Louisiana, and -16.22 in Alaska, exhibiting the opposite impacts of walking mobility on COVID-19 control. The coefficients of RAD indexes were -22.73 in Virginia and -14.88 in Illinois, conveying that intensified public awareness of COVID-19 was the most pivotal NPIs for curving the COVID-19 spread in Virginia and Illinois in the full-blown period.

The results also showed that in Wisconsin, the stay-at-home policies consistently emerged as the most significant NPIs affecting COVID-19 spread throughout the entire year. The coefficients of SI were 26.66, -23.59, and -27.44 in three pandemic periods, respectively. The impacts were positive in the initial outbreak period, and switched to negative in the following periods, depicting that the effectiveness of stay-at-home policies on COVID-19 control began in the rapid spread phase, and enhanced in the full-blown period in Wisconsin. Additionally, in Ohio, Colorado, Wyoming, Maine, South Dakota, Indiana, Hawaii, Kansas, Nebraska, and Arkansas, the COVID-19 spread was primarily shaped by the human mobility intensity in the initial outbreak period, and stay-at-home policies stand out as the most critical factor for COVID-19 control in the rapid spread and full-blown periods. 

\section{Discussion}
\subsection{Significant Implications}

This study has several significant implications. Firstly, this study conducted a demographic-adjusted evaluation of two types of NPIs encompassing public awareness (as measured by the RAD index) and sentiments toward COVID-19 (captured by the N-SAD and P-SAD indexes). The NPIs data set for COVID-19 in the United States at the state level for the year 2020 has been readily accessible via a dedicated GitHub repository ( \url{https://github.com/yimindai0521/Replication_MBSTS_TL}). This comprehensive NPIs data set serves as a fundamental resource for prospective investigations pertaining to societal resilience, ethical considerations within the domain of public health interventions, and the development of adaptable strategies for managing global health crises.

Secondly, this study designed the MBSTS-TL model and applied it to address the high-dimensional challenges in spatiotemporal modelings using the COVID-19 spread as a case study. The MBSTS-TL model offers three distinctive advantages. The model incorporates considerations of spatial dependency and the impact of time lags when examining the relationships between various factors and the target time series. It also leverages the benefits of feature selection to estimate associations within high-dimensional time series. Furthermore, the model effectively addresses concerns related to overfitting when dealing with complex relationships. The MBSTS-TL model serves as a robust analytical instrument for conducting scenario analyses, enabling the evaluation of diverse intervention strategies and their potential consequences on pandemic outcomes. Owing to these inherent advantages, it is also well-suited for elucidating the intricate interplay among societal, economic, environmental, and other public health factors across a wide spectrum of contexts.

Finally, this study unveiled a dynamic pattern in the NPIs influencing the spread of COVID-19 over its various phases. The initial outbreak phase was predominantly driven by human mobility. During the subsequent rapid spread phase, it became evident that the stringency of policies assumed a pivotal role alongside mobility in shaping the pandemic's trajectory. As the pandemic entered its full-blown phase, the significance of public awareness regarding COVID-19 in influencing its progression emerged. This observed pattern underscores the critical importance of adopting a multifaceted strategy that incorporates measures related to mobility constraints, policy stringency, and the implementation of robust public awareness campaigns. The findings of ever-changing determinants that underlie pandemic propagation hold valuable implications for future pandemic preparedness, emphasizing the necessity for a phased and adaptive approach to intervention strategies.


\subsection{Limitations}
While the proposed framework in this study effectively addresses a majority of challenges in estimating the compounding impacts of NPIs on pandemic health outcomes, it is important to acknowledge several limitations that warrant further investigation. 

Accurate estimates of COVID-19 case counts and NPIs are crucial for modeling and understanding the relationship between human responses and COVID-19. However, data uncertainty poses significant challenges in capturing NPIs and the spread of the virus. In the U.S., COVID-19 confirmed case counts were underestimated, primarily due to limited test availability and imperfect test sensitivity, especially during early 2020. \cite{wu2020substantial} pointed out that a substantial number of mild or asymptomatic infections in the U.S. may have gone undetected, as the U.S. Centers for Disease Control and Prevention (CDC) prioritized testing hospitalized patients who tend to exhibit moderate to severe symptoms. Meanwhile, COVID-19 tests based on nasopharyngeal and throat swabs may produce false negative results, leading to the underestimation of COVID-19 cases. To achieve a more accurate understanding of the relationship between human responses and COVID-19 health impacts, a more realistic tracking and assessment of COVID-19 cases is necessary. Uncertainty also exists in the Apple human mobility data, which only records the movement of people using Apple Maps and does not provide a comprehensive representation of overall human mobility. Movements of individuals without GPS-enabled devices or those using different mapping apps cannot be captured in this data. To enhance mobility tracking, it is imperative to incorporate additional human mobility data sources, such as Google human mobility data and SafeGraph data sets.

The concept of "scales of analysis" has long been a geospatial matter that has not yet been systematically elucidated. This term alludes to the level or perspective at which a problem or issue is examined or addressed, encompassing a spectrum from the global level down to the individual level \citep{watson1978scale}. The choice of analysis scale is contingent upon the scale of the problems or issues under consideration. While the COVID-19 pandemic unfolds at the global level, the transmission of the virus is intricately connected to individual interactions. The NPIs and the spread of COVID-19 highlight spatial disparities across various countries, states, counties, and even communities. Consequently, a finer-scale analysis may be more appropriate for investigating the impact of NPIs on the spread of COVID-19. However, due to limitations in the availability of quantitative data concerning COVID-19-related policies, this study's analysis is confined to the state level, treating each state as a single entity and thereby overlooking the spatial heterogeneity in the effects of NPIs on COVID-19. Integrating additional social sensing data to track human activities at a finer granularity enhances the depth and precision of our comprehension of the human-pandemic dynamic system, providing a more comprehensive insight into reality.

Moreover, the influence of NPIs on the dissemination of COVID-19 is contingent upon various localized factors, such as population density, the influx of individuals from initial outbreak epicenters, and prevailing mobility patterns. Future studies considering these factors could have furnished a more comprehensive and nuanced understanding of how NPIs impact the transmission dynamics of COVID-19.

\section{Conclusion}

This study collected Twitter data, COVID-19 case rates, Apple human mobility data, and the stringency of stay-at-home policies in the United States during the year 2020. The overarching objectives of this investigation encompassed two fundamental research lacunae. Initially, it aspired to elucidate the spatiotemporal disparities in NPIs, encompassing aspects such as public awareness, sentiment towards COVID-19, human mobility patterns, and the rigor of COVID-19 policies, within the United States throughout 2020. To this end, the study introduced the RAD index to estimate demographically adjusted public awareness towards COVID-19 using Twitter data. Furthermore, it produced weekly N-SAD and P-SAD indices at the state level, quantitatively capturing negative and positive public sentiment regarding COVID-19 based on Twitter data. Secondly, it aimed to design and employ a statistical machine learning model for the comprehensive modeling of the cumulative effects of NPIs on COVID-19 health outcomes, with a specific emphasis on accounting for spatial interdependencies and temporal lag effects in the relationships. The MBSTS-TL model was proposed to uncover interconnected relationships between NPIs and COVID-19 health outcomes while considering spatial dependencies and the time lag effects of NPIs on the transmission of COVID-19.

The outcomes of this research have yielded significant insights. First, it has unveiled spatiotemporal discrepancies in NPIs pertaining to COVID-19 at the state level in the United States. Nationally, during the initial outbreak period (from week 9 to week 22, spanning from late February to the end of May), public awareness experienced a rapid increase, reaching a significant peak. Sentiment towards COVID-19 exhibited its most negative trends during this period compared to the remainder of 2020. The stringency of stay-at-home policies also increased rapidly and was maintained at a relatively stringent level. Human mobility, whether by walking, driving, or transit, witnessed varying degrees of decline and recovery. At the state level, NPIs displayed temporal and geographical variations. For instance, Louisiana consistently exhibited lower levels of public awareness about the pandemic among Twitter users throughout 2020, while Twitter users in Vermont consistently demonstrated a more optimistic outlook regarding the pandemic, and those in Wyoming generally expressed a more pessimistic sentiment.

Second, the developed MBSTS-TL model demonstrated high accuracy during the initial outbreak and rapid spread periods, with consistently low normalized absolute errors. The findings of the MBSTS-TL model indicated a shift in the determinants of COVID-19 health impacts over time, transitioning from an emphasis on human mobility during the initial outbreak period to a combination of human mobility and stay-at-home policies during the rapid spread period, and eventually involving human mobility, stay-at-home policies, and public awareness towards COVID-19 in the full-blown phase.

The NPIs data set and the proposed MBSTS-TL model offer valuable insights for diverse applications. The NPIs data set provides a valuable resource for social scientists aiming to comprehend the dynamics of human activities across space and time during the pandemic. Additionally, the MBSTS-TL model can be effectively employed to unveil complex interrelationships within human-public health systems and human-environment dynamics. The results serve to identify the evolving determinants of NPIs on pandemic spread, thereby offering guidance to policymakers for the implementation of phased and adaptive strategies aimed at mitigating the adverse impacts of future pandemics.

\section*{Acknowledgements}
This study is partially supported by the Data Resource Develop Program Award from the Texas A\&M Institute of Data Science (TAMIDS) and the Seed Fund Award from the College of Arts \& Sciences at Texas A\&M University. Any opinions, findings, conclusions, or recommendations expressed in this material are those of the authors and do not necessarily reflect the views of the funding agencies.

\section*{Data availability statement}

The data used in this research were derived from the following resources available in the public domain: Twitter Application Programming Interface (API) for Academic Research (\url{https://developer.twitter.com/en/products/twitter-api/academic-research}),  Oxford Covid-19 Government Response Tracker (OxCGRT) (\url{https://github.com/OxCGRT/covid-policy-tracker#oxford-covid-19-government-response-tracker-oxcgrt}), and COVID-19 Data Repository by the Center for Systems Science and Engineering (CSSE) at Johns Hopkins University (\url{https://github.com/CSSEGISandData/COVID-19}).  The RAD, N-SAD, and P-SAD dataset, as well as the Replication code and mbsts-tl function generated in this study, are available as a GitHub repository ( \url{https://github.com/yimindai0521/Replication_MBSTS_TL}).





\bibliographystyle{ecta}
\small{
\bibliography{ref}
}


\end{document}